\documentstyle[12pt]{article}
\begin{document}
\begin{titlepage}
\begin{center}

\null
\vskip-1truecm
\rightline{IC/93/413}
\vskip1truecm
{International Atomic Energy Agency\\
and\\
United Nations Educational Scientific and Cultural Organization\\
\medskip
INTERNATIONAL CENTRE FOR THEORETICAL PHYSICS\\}
\vskip2truecm
{\bf A (p, q) DEFORMATION\\
OF THE UNIVERSAL ENVELOPING\\ SUPERALGEBRA U(osp(2/2))\\}
\vskip2truecm
{Preeti Parashar\\
International Centre for Theoretical Physics, Trieste, Italy.\\}
\end{center}
\vskip1truecm
\centerline{ABSTRACT}
\bigskip

We investigate a two--parameter quantum deformation of the universal
enveloping orthosymplectic superalgebra U(osp(2/2)) by extending the
Faddeev--Reshetikhin--Takhtajan formalism to the supersymmetric case. It is
shown that U$_{p,q}$(osp(2/2)) possesses a non--commutative, non--cocommutative
Hopf algebra structure. All the results are expressed in the standard form
using quantum Chevalley basis.

\vskip2truecm

\begin{center}
{MIRAMARE -- TRIESTE\\
\medskip
December 1993\\}
\end{center}

\end{titlepage}

There has been an enormous interest in quantum deformations of Lie groups and
Lie algebras during the last couple of years. The next step in this direction
has been to extend these ideas to supergroups and superalgebras [1--6]. The aim
of the present investigation is to obtain a two--parameter deformation of the
universal enveloping algebra of the orthosymplectic Lie superalgebra osp(2/2).
This is achieved by extending the basic considerations of the $R$--matrix
approach proposed by the Leningrad school [7], to the supersymmetric case. We
also find out the action of various maps on the generators of
U$_{p,q}$(osp(2/2)) to show that it is indeed equipped with a Hopf algebra [8]
structure which is non--commutative as well as non--cocommutative. Finally, all
the super--commutation relations obtained in the $R$--matrix framework are
transformed into the standard form with the help of the quantum analogous of
the Chevalley generators.

The algebra of functions on a quantum supergroup is defined  by the relations
\begin{equation}
\hat R\ T_1T'_2=T_1T'_2\ \hat R
\end{equation}
where the matrix $\hat R$ is a solution of braid QYBE and corresponds to that
particular quantum supergroup. $T$ is the transformation matrix and
\begin{eqnarray}
(T_1)^{ab}_{cd} & = & (T\otimes I)^{ab}_{cd}=(-1)^{c(b+d)}\ T^a_c\ \delta^b_d
\nonumber \\
(T_2)^{ab}_{cd} & = & (I\otimes T)^{ab}_{cd}=(-1)^{a(b+d)}\ T^b_d\ \delta^a_c
\\
T'_2 & = & {\cal P}\ T_1\ {\cal P} \nonumber
\end{eqnarray}
where ${\cal P}$ is the super--permutation matrix
\begin{equation}
({\cal P})^{ab}_{cd}=(-1)^{ab}\ \delta^a_d\delta^b_c\ .
\end{equation}

However, for the case of orthosymplectic supergroups we need to impose
additional conditions:
\begin{equation}
CT^{st}\ C^{-1}T=TCT^{st}\ C^{-1}=I\ .
\end{equation}
Here $C$ is an antidiagonal metric and $T^{st}$ is the supertranspose of $T$
defined as
\begin{equation}
(T^i_j)^{st}=(-1)^{j(i+j)}\ T^j_i\ .
\end{equation}
Relation (4) is the modified form of the conditions given by FRT [7] for
orthogonal and symplectic groups separately. So the algebra of functions on the
orthosymplectic group is defined by relations (1) and (4) together.

Let us now consider a specific example of the quantum supergoup OSp(2/2). The
non--vanishing elements of the $(16\times 16)\ \hat R$ matrix are [9]:
\begin{eqnarray}
\hat R^{11}_{11} & = & \hat R^{44}_{44}=-(pq)^{-1/2},\quad
\hat R^{22}_{22}=\hat R^{33}_{33}=(pq)^{1/2},\nonumber \\
\hat R^{12}_{21} & = & \hat R^{31}_{13}=\hat R^{24}_{42}=\hat R^{43}_{34}
=(p/q)^{1/2},\nonumber \\
\hat R^{21}_{12} & = & \hat R^{13}_{31} =\hat R^{42}_{24}=\hat R^{34}_{43}
=(p/q)^{-1/2},\nonumber \\
\hat R^{12}_{12} & = &\hat R^{13}_{13}=\hat R^{24}_{24}=\hat R^{34}_{34}
=(pq)^{1/2}-(pq)^{-1/2},\nonumber \\
\hat R^{32}_{23} & = & \hat R^{23}_{32}=(pq)^{-1/2},\quad
\hat R^{41}_{14}=\hat R^{14}_{41}=-(pq)^{1/2},\nonumber \\
\hat R^{14}_{14} & = & \left( (pq)^{1/2}-(pq)^{-1/2}\right)(1-(pq)^{-1}),\ \
\hat R^{23}_{23}=\left( (pq)^{1/2}-(pq)^{-1/2}\right) (1+(pq)^{-1}), \nonumber
\\
\hat R^{23}_{14} & = & \hat
R^{14}_{23}=i(pq)^{-1}\left( (pq)^{1/2}-(pq)^{-1/2}\right),\ \  \hat
R^{14}_{32}=\hat R^{32}_{14}=-i\left( (pq)^{1/2}-(pq)^{-1/2}\right)\ .
\end{eqnarray}
where $p$ and $q$ are two (complex) deformation parameters. $T\equiv
(T^i_j)_{i,j=1,\dots 4}$ is a $4\times 4$ super--matrix acting on a quantum
vector space generated by the elements $x^1,x^2,x^3,x^4$, where the indices
(1,4) are fermionic and (2,3) are bosonic. The metric $C$ in this case is
$4\times 4$ and given by [9]
\begin{equation}
C=\left(\matrix{
0&0&0&
i(pq)^{-1/2}\cr
0&0&(pq)^{-1/2}&0\cr
0&-(pq)^{1/2}&0&0\cr
i(pq)^{1/2}&0&0&0\cr}
\right)\ .
\end{equation}
A crucial feature of quantum OSp(2/2) is that it leaves the following
constraint invariant:
\begin{equation}
 x^tCx=i(pq)^{1/2}
x^4x^1-(pq)^{1/2}x^3x^2+(pq)^{-1/2}x^2x^3+i(pq)^{-1/2}x^1x^4=0\ .
\end{equation}
The homogeneous quadratic part of the algebra is preserved in the usual way by
virtue of the (super--) RTT relns. (1) whereas the inhomogeneous part is left
covariant by virtue of (4).

Thus we obtain the complete set of deformed commutation relations among the
generators $T^i_j$ of the functions on the quantum supergroup OSp$_{p,q}$(2/2),
by substituting the matrices $\hat R,T$ and $C$ in (1) and (4). It turns out
that the elements $(T^i_j)_{ij=1,\dots 4},\ i\not= j'$ (where $1'=4,2'=3$)
behave
as odd generators (since their squares vanish) and the rest as even.

Our main purpose in this letter is to obtain a deformation of the dual object
i.e. the universal enveloping superalgebra U\ osp(2/2). The generators are
arranged in $4\times 4$ upper and lower triangular matrices $L^{(\pm
)}\equiv (L^{(\pm )}{i\atop j})_{i,j=1,\dots 4}$ of regular functionals and
obey the (super--) commutation relations
\begin{equation}
R_{21}\ L^{(\varepsilon_1)}_1\ L^{(\varepsilon_2)}_2 =
L^{(\varepsilon_2)}_2\ L^{(\varepsilon_1)}_1\ R_{21}
\end{equation}
where $(\varepsilon_1,\varepsilon_2)=(+,+), (+-),(-,-), R_{21}={\cal
P}R_{12}{\cal P}$. For the sake of convenience we recast the above relations in
terms of $\hat R(={\cal P}R)$
\begin{equation}
\hat R_{12}\ {\cal
P}_{12}\ \eta_{12}L^{(\varepsilon_1)}_1\ \eta_{12}\ L^{(\varepsilon_2)}_2=
L^{(\varepsilon_2)}_2\ \eta_{12}\ L^{(\varepsilon_1)}_1\ \eta_{12}\ \hat
R_{12}\ {\cal P}_{12}
\end{equation}
where $(\eta_{12}){ab\atop cd}=(-1)^{ab}\ \delta^a_c\delta^b_d$ is a diagonal
phase factor. In (10) $\hat R_{12}$ is graded and the tensor products $L_1$ and
$L_2$ remain ungraded (i.e. without any phase factors), contrary to (9) in
which $R_{12}$ matrix is ungraded (i.e. $R_{21}$ is graded) and $L_1$ and $L_2$
are graded and are defined in the same fashion as $T_1$ and $T_2$.

The pairing $\langle T^i_j,L^{(+)}{k\atop \ell}\rangle = R^{ik}_{j\ell},\langle
T^i_j,L^{(-)}{k\atop\ell}\rangle = R^{(-1)}{ki\atop\ell j}$ puts a
restriction on
the number of independent generators. We shall list down here the commutation
relations (obtained from (10)) involving only the independent generators
$L^{(+)}{1\atop 2},L^{(+)}{2\atop 3},L^{(-)}{2\atop 1},L^{(-)}{3\atop 2}$
and the
diagonal elements $L^{(+)}{a\atop a}$.
\begin{eqnarray}
\left[ L^{(\pm )}{\scriptstyle{a\atop a}}, L^{(\pm )}
{\scriptstyle{b\atop b}}\right] & = & 0\quad a,b=1,2,3,4, \nonumber \\
L^{(+)}{\scriptstyle{({a\atop b})\atop
({a\atop b})}}\ L^{(+)}{\scriptstyle{1\atop 2}} & = &
(p)^{({+1\atop -1})}\ L^{(+)}{\scriptstyle{1\atop 2}}\
L^{(+)}\scriptstyle{{({a\atop b})\atop ({a\atop b})}},\ \  a=1,2,\ \ b=3,4,
\nonumber \\
 L^{(-)}{\scriptstyle{({a\atop b})\atop ({a\atop b})}}\ L^{(+)}
{\scriptstyle{1\atop 2}} & = & (q)^{({-1\atop +1})}\ L^{(+)}
{\scriptstyle{1\atop 2}}\ L^{(-)}
\scriptstyle{{({a\atop b})\atop ({a\atop b})}},\nonumber \\
L^{(+)}
{\scriptstyle{({a\atop b})\atop ({a\atop b})}}\ L^{(-)}{\scriptstyle{2\atop 1}}
& = & (p)^{({-1\atop +1})}\ L^{(-1)} {\scriptstyle{2\atop 1}}\ L^{(+)}
{\scriptstyle{({a\atop
b})\atop ({a\atop b})}}, \nonumber \\
L^{(-)}{\scriptstyle{({a\atop b})\atop ({a\atop b})}}\
L^{(-)}{\scriptstyle{2\atop 1}} & = & (q)^{({+1\atop -1})}\ L^{(-)}
{\scriptstyle{2\atop 1}}\ L^{(-)}
{\scriptstyle{({a\atop b})\atop ({a\atop b})}}, \nonumber \\
L^{(\pm )}{\scriptstyle{({1\atop 4})\atop ({1\atop 4})}}\
L^{(+)}{\scriptstyle{2\atop 3}} & = & \left({q\over p}\right)^{\left({+1\atop
-1}\right)}\ L^{(+)}{\scriptstyle{2\atop 3}}\ L^{(\pm )}
{\scriptstyle{({1\atop 4})\atop
({1\atop 4})}}, \nonumber \\
L^{(\pm )}{\scriptstyle{({2\atop 3})\atop ({2\atop 3})}}\
L^{(+)}{\scriptstyle{2\atop 3}} & = & (pq)^{\left({-1\atop +1}\right)}\
L^{(+)}{\scriptstyle{2\atop 3}}\ L^{(\pm )}
{\scriptstyle{({2\atop 3})\atop ({2\atop 3})}},
\nonumber \\
L^{(\pm )}{\scriptstyle{({1\atop 4})\atop ({1\atop 4})}}\
L^{(-)}{\scriptstyle{3\atop 2}} & = & \left({q\over p}\right)^{\left({-1\atop
+1}\right)}\ L^{(-)} {\scriptstyle{3\atop 2}}\ L^{(\pm
)}{\scriptstyle{({1\atop 4})\atop ({1\atop 4})}},
\nonumber \\
L^{(\pm )}{\scriptstyle{({2\atop 3})\atop ({2\atop 3})}}\ L^{(-)}
{\scriptstyle{3\atop 2}} & = &
(pq)^{\left({+1\atop -1}\right)}\ L^{(-)}{\scriptstyle{3\atop 2}}\ L^{(\pm
)}{\scriptstyle{({2\atop 3})\atop ({2\atop 3})}},
\nonumber \\
\left( L^{(+)}{\scriptstyle{1\atop 2}}\right)^2=0 & = &
\left( L^{(-)}{\scriptstyle{2\atop
1}}\right)^2,\nonumber\\
\left\{ L^{(+)}{\scriptstyle{1\atop 2}},L^{(-)}
{\scriptstyle{2\atop 1}}\right\}_{p/q} & = &
\left( (pq)^{1/2}-(pq)^{-1/2}\right)\left( L^{(-)}
{\scriptstyle{2\atop 2}}\ L^{(+)}{\scriptstyle{1\atop 1}}-
L^{(+)}{\scriptstyle{2\atop 2}}\ L^{(-)}
{\scriptstyle{1\atop 1}}\right),\nonumber \\
\left[ L^{(+)}{\scriptstyle{2\atop 3}}, L^{(-)}{\scriptstyle{3\atop 2}}\right]
&
= & \left( (pq)- (pq)^{-1}\right)\left( L^{(-)}{\scriptstyle{3\atop 3}}\
L^{(+)}{\scriptstyle{2\atop 2}} - L^{(+)}{\scriptstyle{3\atop 3}}\
L^{(-)}{\scriptstyle{2\atop 2}}\right)\ ,
\end{eqnarray}
where $L^{(\pm )}{({1\atop 4})\atop ({1\atop 4})}\ L^{(+)}{2\atop 3}
=\left({q\over p}\right)^{\left({+1\atop -1}\right)}\ L^{(+)}{2\atop 3}\
L^{(\pm )}{({1\atop 4})\atop ({1\atop 4})}$ implies $L^{(\pm
)}{1\atop 1}\ L^{(+)}{2\atop 3}={q\over p} \ L^{(+)}{2\atop 3}\ L^{(\pm )}
{1\atop 1}$ and $L^{(\pm )}{4\atop 4}\ L^{(+)}{2\atop 3} =\left({q\over
p}\right)^{-1}L^{(+)}{2\atop 3}L^{(\pm )}{4\atop 4}$ and so on. The
anticommutator $\{ ,\}_{p/q}$ is defined as [4]
\begin{equation}
\{ A,B\}_{p/q}=\left({p\over q}\right)^{1/2}AB+\left({p\over
q}\right)^{-1/2}BA\ .
\end{equation}
These generators play the role of quantum analogue of the Cartan--Weyl basis.
We notice that $L^{(+)}{1\atop 2}$ and $L^{(-)}{2\atop 1}$ are fermionic in
nature whereas $L^{(+)}{2\atop 3}$ and $L^{(-)}{3\atop 2}$ are bosonic.

The above set of deformed (super--) commutation relations (11) are supplemented
by the conditions
\begin{equation}
L^{(\pm )}\ C^{st}\ L^{(\pm )st}(C^{-1})^{st}=
C^{st}\ L^{(\pm )st} (C^{-1})^{st}\ L^{(\pm )} =I
\end{equation}
which yield
\begin{eqnarray}
L^{(+)}{\scriptstyle{1\atop 1}}\ L^{(+)}{\scriptstyle{4\atop 4}} &=& 1  =
L^{(-)}{\scriptstyle{1\atop 1}}\  L^{(-)}{\scriptstyle{4\atop 4}}, \nonumber \\
L^{(+)}{\scriptstyle{2\atop 2}}\ L^{(+)}{\scriptstyle{3\atop 3}} &=& 1=
L^{(-)}{\scriptstyle{2\atop 2}}\ L^{(-)}{\scriptstyle{3\atop 3}}\ .
\end{eqnarray}
(13) can be regarded as the dual of (4). It is transparent that the
relations (14) belong to the centre of the algebra.

We shall now give the action of the structure maps on the generators. The
coproduct $\Delta (L^{(\pm )})=L^{(\pm )}\dot{\otimes} L^{(\pm )}$ is given by

\newpage

\begin{eqnarray}
\Delta \left (L^{(\pm )}{\scriptstyle{a\atop a}}\right) & = &
L^{(\pm )}{\scriptstyle{a\atop a}}\otimes L^{(\pm )}
{\scriptstyle{a\atop a}}\quad a=1,\dots ,4,\nonumber \\
\Delta\left( L^{(+)}{\scriptstyle{1\atop 2}}\right) & = & L^{(+)}
{\scriptstyle{1\atop
1}}\otimes L^{(+)}{\scriptstyle{1\atop 2}}+L^{(+)}
{\scriptstyle{1\atop 2}}\otimes
L^{(+)}{\scriptstyle{2\atop 2}}\nonumber \\
\Delta\left( L^{(+)}{\scriptstyle{2\atop
3}}\right) & = & L^{(+)}{\scriptstyle{2\atop 2}}\otimes L^{(+)}
{\scriptstyle{2\atop
3}}+L^{(+)}{\scriptstyle{2\atop 3}}\otimes L^{(+)}
{\scriptstyle{3\atop 3}}\nonumber \\
\Delta\left( L^{(-)}{\scriptstyle{2\atop 1}}\right) & = &
 L^{(-)}{\scriptstyle{2\atop
1}}\otimes L^{(-)}{\scriptstyle{1\atop 1}}+L^{(-)}
{\scriptstyle{2\atop 2}}\otimes
L^{(-)}{\scriptstyle{2\atop 1}}\nonumber \\
\Delta\left( L^{(-)}{\scriptstyle{3\atop
2}}\right) & = & L^{(-)}{\scriptstyle{3\atop 2}}\otimes
L^{(-)}{\scriptstyle{2\atop 2}}+L^{(-)}{\scriptstyle{3\atop 3}}\otimes
L^{(-)}{\scriptstyle{3\atop 2}}\ .
\end{eqnarray}
The co--unit $\varepsilon \left( L^{(\pm )}{i\atop j}\right)$
yields \begin{equation} \varepsilon\left( L^{(\pm )}
\scriptstyle{{a\atop a}}\right) =1\quad
{\rm and\ zero\ for\ all\ others}.
\end{equation}
The antipode $S$ for the orthosymplectic superalgebras can be defined as
\begin{equation}
S(L^{(\pm )})=C^{st}\ L^{(\pm )st}(C^{-1})^{st}
\end{equation}
which follows from (13). This gives
\begin{eqnarray}
S\left( L^{(\pm )}{\scriptstyle{1\atop 1}}\right) & = & L^{(\pm )}
{\scriptstyle{4\atop 4}},\
S\left( L^{(+)} {\scriptstyle{1\atop 2}}\right)=iL^{(+)}
{\scriptstyle{3\atop 4}},\  S\left(
L^{(+)}{\scriptstyle{2\atop 3}}\right) = - pq\ L^{(+)}
{\scriptstyle{2\atop 3}},\nonumber \\
S\left( L^{(\pm )}{\scriptstyle{2\atop 2}}\right) & = & L^{(\pm )}
{\scriptstyle{3\atop 3}},\
S\left( L^{(-)}{\scriptstyle{2\atop 1}}\right) =iL^{(-)}
{\scriptstyle{4\atop 3}},\  S\left(
L^{(-)}{\scriptstyle{3\atop 2}} \right) =-(pq)^{-1}\ L^{(-)}
{\scriptstyle{3\atop 2}}\ .
\end{eqnarray}
The above maps endow the quantum Lie superalgebra with a non--commutative and
non--cocommutative Hopf (super) algebra structure. We have thus obtained a
two--parameter quantum universal enveloping superalgebra U$_{p,q}$(osp(2/2)).

The essential task now is to express these results of the $R$--matrix approach
into some standard known form so as to make contact with the classical case.
The Lie superalgebra osp(2/2) [10] is of rank 2 and has 6 generators
$(H_i,x^+_i,x^-_i, i=1,2)$ corresponding to the two simple roots. We make the
following identification between the two sets of generators:
\begin{eqnarray}
L^{(+)}{\scriptstyle{1\atop 2}}\left( L^{(-)}{\scriptstyle{1\atop 1}}\
L^{(-)}{\scriptstyle{2\atop 2}}\right)^{1/2} & = & \left(
(pq)^{1/2}-(pq)^{-1/2}\right)\ x^+_1 \nonumber \\ L^{(-)}
{\scriptstyle{2\atop 1}}\left(
L^{(-)}{\scriptstyle{1\atop 1}}\ L^{(-)}{\scriptstyle{2\atop 2}}\right)^{-1/2}
&
= & -\left( (pq)^{1/2}-(pq)^{-1/2}\right)\ x^-_1 \nonumber \\
L^{(+)}{\scriptstyle{2\atop 3}} &
= & \left( (pq)-(pq)^{-1}\right)\ x^+_2 \nonumber\\
 L^{(-)}{\scriptstyle{3\atop 2}} & = &
-\left( (pq)-(pq)^{-1}\right)\ x^-_2\ .
\end{eqnarray}
If we define the logarithms $H_1,H_2$ by
\begin{eqnarray}
L^{(+)}{\scriptstyle{1\atop 1}} & = & q^{-H_1/2-H_2}\ p^{-H_1/2},\ \
L^{(-)}{\scriptstyle{1\atop 1}} = q^{H_1/2}\ P^{H_1/2+H_2}, \nonumber \\
L^{(+)}{\scriptstyle{2\atop 2}} & = & p^{-H_2},\ \  L^{(-)}
{\scriptstyle{2\atop 2}} = q^{H_2},
\ \  L^{(+)}{\scriptstyle{3\atop 3}}=p^{H_2},\ \ L^{(-)}
{\scriptstyle{3\atop 3}}=q^{-H_2}
\end{eqnarray}
where
\begin{equation}
q=e^h,\quad p=e^{h'}
\end{equation}
then we obtain the following commutation relations between the new set of
generators:

\newpage

\begin{eqnarray}
[H_1,x^+_1] & = & 2\left({h'-h\over h+h'}\right)\ x^+_1, \nonumber \cr
[H_1,x^-_1] & = & 2\left({h-h'\over h+h'}\right)\ x^-_1, \nonumber \cr
[H_1,x^+_2] & = & -4{h'\over h+h'}\ x^+_2=2a_{12}\left({h'\over h+h'}\right)\
x^+_2,  \nonumber \cr
[H_1,x^-_2] & = & 4{h'\over h+h'}\
x^-_2=-2a_{12}\left({h'\over h+h'}\right)\ x^-_2,  \nonumber \cr
[H_2,x^+_1] & = & -x^+_1=a_{21}\ x^+_1, \nonumber \cr
[H_2,x^-_1] & = & +x^-_1=-a_{21}\ x^-_1, \nonumber \cr
[H_2,x^+_2] & = &
2x^+_2=a_{22}\ x^+_2, \nonumber \cr
[H_2,x^-_2] & = &
-2x^-_2=-a_{22}\ x^-_2, \nonumber \cr [H_1,H_2] & = & 0,\nonumber \cr
\{ x^+_1,x^-_1\}_{p/q} & = & {(pq)^{H_1/2}-(pq)^{-H_1/2}\over
(pq)^{1/2}-(pq)^{-1/2}}\equiv [H_1]_{(pq)^{1/2}} \nonumber \cr
[x^+_2,x^-_2] & =
& {(pq)^{H_2}-(pq)^{-H_2}\over (pq)-(pq)^{-1}}\equiv [H_2]_{pq}
\end{eqnarray}
\begin{equation}
\end{equation}
where ${\cal A}(a_{ij})$ is the Cartan matrix for osp(2/2) given by
\begin{equation}
{\cal A}=\left(\matrix{ \ \ 0&-2\cr -1&\ \ 2\cr}\right)
\end{equation}
which is symmetrized by the matrix ${\cal D}=diag(1,2)$. These generators
$(H_i,x^\pm_i)$ act as $q$--analogues of Chevalley basis. It is however more
convenient to introduce the elements
\begin{equation}
k_1=(pq)^{H_1/2},\quad k_2=(pq)^{H_2}\ .
\end{equation}
Now the super--commutation relations can be written down explicitly as:
\begin{eqnarray}
k_1\ x^\pm_1 & = & \left({p\over q}\right)^{\left({+1\atop -1}\right)}\
x^\pm_1\ k_1,\ \ k_2\ x^\pm_1 = (pq)^{\left({-1\atop +1}\right)}\
x^\pm_1\ k_2, \nonumber \\
k_1\ x^\pm_2 & = & (p)^{\left({-2\atop +2}\right)}\ x^\pm_2\
k_1,\ \ k_2\ x^\pm_2 = (pq)^{\left({+2\atop -2}\right)}\ x^\pm_2\ k_2,
\nonumber
\\
k_1\ k_2 & = & k_2\ k_1, \nonumber \\
\{ x^+_1,x^-_1\}_{p/q} & = & {k_1-k^{-1}_1\over (pq)^{1/2}-(pq)^{-1/2}},\ \
[x^+_2,x^-_2]={k_2-k^{-1}_2\over (pq)-(pq)^{-1}}\ .
\end{eqnarray}
It is also useful to introduce additional generators $x^+_3,x^-_3,k_3$ (in
complete analogy with the Lie superalgebra) using the $q$--adjoint operation
[11]:
\begin{eqnarray}
x^+_3 & \equiv & [x^+_2,x^+_1]=q\ x^+_2\ x^+_1-q^{-1}\ x^+_1\ x^+_2,
\nonumber \\
x^-_3 & \equiv & [x^-_1,x^-_2] =q^{-1}\ x^-_1\ x^-_2-p\ x^-_2\ x^-_1\ .
\end{eqnarray}
Then one obtains
\begin{equation}
\{ x^-_3,x^+_3\}_{p/q}={\left({q\over p}\right)^{H_2}(k_3-k^{-1}_3)\over
(pq)^{1/2}-(pq)^{-1/2}},\quad {\rm where}\quad k_3=k_1k_2
\end{equation}
and the quantum analogue of the Serre relations
\begin{eqnarray}
[x^+_2,x^+_3] & = & 0,\quad [x^-_2,x^-_3]=0, \cr
[x^+_1,[x^+_1,x^+_3]] &= & 0,\quad [x^-_1,[x^-_1,x^-_3]]=0\ .
\end{eqnarray}
Notice the strangeness in the form of $x^+_3$ which involves only one
deformation parameter and $x^-_3$ which has both $p$ and $q$.

We have thus arrived at a two--parameter deformation of the universal
enveloping algebra U\ osp(2/2). When $p=q\not= 1$, this coincides with the
results obtained in [5, 6] for a single parameter, and in the limit $p=q=1$ we
recover the classical Lie superalgebra osp(2/2).

The expressions for the coproduct, co--unit and antipode for the new generators
assume the following form.
\medskip

\noindent{\em Co--product:}

\begin{eqnarray}
\Delta (k_i) & = & k_i\otimes k_i\qquad i=1,2,3, \nonumber \\
\Delta (x^+_1) & = & x^+_1\otimes k^{1/2}_1\left({q\over p}\right)^{H_2/2}+
k^{-1/2}_1\left({q\over p}\right)^{-H_2/2}\otimes x^+_1 \nonumber \\
\Delta (x^-_1) & = & x^-_1\otimes k^{1/2}_1\left({q\over p}\right)^{-H_2/2}+
k^{-1/2}_1\left({q\over p}\right)^{H_2/2}\otimes x^-_1 \nonumber \\
\Delta (x^+_2) & = & x^+_2\otimes k_2\ q^{-H_2}+
k^{-1}_2\ q^{H_2}\otimes x^+_2 \nonumber \\
\Delta (x^-_2) & = & x^-_2\otimes k_2\ p^{-H_2}+ k^{-1}_2\ p^{H_2}\otimes x^-_2
\nonumber \\
\Delta (x^+_3) & = & k^{-1/2}_3\otimes x^+_3+x^+_1p^{-H_2}\otimes
x^+_2\ k_2^{1/2}\left({q\over p}\right)^{H_2/2}+x^+_3\otimes k^{1/2}_3
\nonumber \\
\Delta (x^-_3) & = & k^{-1/2}_3\left({q\over p}\right)^{H_2}\otimes x^-_3 +
x^-_2\ k^{-1/2}_1\left({q\over p}\right)^{3H_2/2}\otimes x^-_1\
q^{H_2}\left({q\over p}\right)^{H_2}+ x^-_3\otimes k^{1/2}_3\left(
{q\over p}\right)^{H_2}\nonumber \\
&&
\end{eqnarray}
\medskip

\noindent {\em Co--unit:}
\begin{equation}
\varepsilon (1)=1,\quad \varepsilon (k_i)=1,\quad \varepsilon (x^\pm_i)=0,
\quad i=1,2,3
\end{equation}
\medskip

\noindent {\em Antipode} (co--inverse):

\begin{eqnarray}
S(k_i) & = & k^{-1}_i,\quad i=1,2,3, \nonumber \\
S(x^+_1) & = & -k^{1/2}_1\left({q\over p}\right)^{H_2/2}\ x^+_1
\left({q\over p}\right)^{-H_2/2}\ k^{-1/2}_1 \nonumber \\
S(x^-_1) & = & -k^{1/2}_1\left({q\over p}\right)^{-H_2/2}\ x^-_1
\left({q\over p}\right)^{H_2/2}\ k^{-1/2}_1 \nonumber \\
S(x^+_2) & = & -k_2\ q^{-H_2}\ x^+_2\ q^{H_2}\ k^{-1}_2 \nonumber \\
S(x^-_2) & = & -k_2\ p^{-H_2}\ x^-_2\ p^{H_2}\ k^{-1}_2 \nonumber \\
S(x^+_3) & = & -k^{1/2}_3 x_3^+k^{-1/2}_3+p^{H_2}k^{1/2}_1\left({q\over
p}\right)^{H_2/2}x^+_1\left({q\over
p}\right)^{-H_2/2}k^{-1/2}_1x^+_2k^{1/2}_1\left({q\over
p}\right)^{H_2/2}k^{-1/2}_3 \nonumber \\
S(x^-_3) & = & -k^{1/2}_3 \left({q\over p}\right)^{-H_2}x^-_3
\left({q\over p}\right)^{-H_2}k^{-1/2}_3+k^{1/2}_1
\left({q\over p}\right)^{-3H_2/2}q^{H_2}x^-_2q^{-H_2}x^-_1q^{H_2}k^{-1/2}_3\
{}.
\end{eqnarray}
We conclude with the following remarks. The results obtained for quantum
supergroup OSp(2/2) and its corresponding universal enveloping algebra
U$_{p,q}$ (osp(2/2)) can be generalized to higher dimensions i.e. to
U(osp(2m/2n)) which would involve many more deformation parameters. The
supersymmetric formulae we have presented here, ((1), (4), (10) and (13)), are
very general and hold for all types of orthosymplectic supergroups and their
superalgebras belonging to the B, C and D series. In the future, we wish to
study the representation theories of these quantum superalgebras.
 \vskip2truecm

\noindent{\large {\bf Acknowledgments}}
\bigskip

The author is grateful to L. Dabrowski, V.K. Dobrev and X--C Song for useful
discussions at various stages of this work and to S. Mukherji for
encouragement. She would like to thank Professor Abdus Salam, the International
Atomic Energy Agency and UNESCO for hospitality at the International Centre for
Theoretical Physics, Trieste.

\newpage

\centerline{REFERENCES}
\bigskip
\begin{description}

\item{1.}
Chaichian M. and Kulish P., 1990, Phys. Lett. {\bf 234B}, 72;\\
Chaichian M., Kulish P. and Lukierski J., 1991, Phys. Lett. {\bf 262B}, 43.

\item{2.}
Kulish P. and Reshetikhin N. Yu., 1989, Lett. Math. Phys. {\bf 18}, 143.

\item{3.}
Schmidke W.B., Vokos S.P. and Zumino B., 1990, Z. Phys. {\bf C48}, 249.

\item{4.}
Dabrowski L. and Wang Lu--yu, 1991, Phys. Lett. {\bf 266B}, 51.

\item{5.}
Liao L. and Song X--C, 1991, Mod. Phys. Lett. {\bf A6}, 959; J. Phys. {\bf
A24}, 5451.

\item{6.}
Floreanini R., Spiridonov V.P. and Vinet L., 1991, Comm. Math. Phys. {\bf 137},
149\\
and references therein.

\item{7.}
Faddeev L.D., Reshetikhin N.Yu. and Takhtajan L.A., LOMI preprint E--14--87;\\
Leningrad Math. J. {\bf 1} (1990) 193.

\item{8.}
Abe E., Hopf Algebras, Cambridge University Press, Cambridge, 1980.

\item{9.}
Parashar P. and Soni S.K., 1993, J. Phys. {\bf A26}, 921.

\item{10.}
Kac. V.G., Adv. Math. {\bf 26} (1977) 8; Comm. Math. Phys. {\bf 53} (1977) 31.

\item{11.}
Rosso M., Comm. Math. Phys. {\bf 124} (1989) 307.
\end{description}

\end{document}